\documentclass[prb,superscriptaddress,showpacs,showkeys,twocolumn,10pt]{revtex4-1}
\usepackage{amsfonts,amsmath,amssymb,graphicx,epstopdf,verbatim,color}
\usepackage[english]{babel}

\begin{document}

\title{Single particle polariton properties in doped quantum well microcavities: role of the Fermi edge singularity and Anderson orthogonality catastrophe }
\author{Maarten Baeten}
\affiliation{Theory Of Quantum and Complex Systems, Universiteit Antwerpen, Universiteitsplein 1, B-2610 Antwerpen, Belgium}
\author{Michiel Wouters}
\affiliation{Theory Of Quantum and Complex Systems, Universiteit Antwerpen, Universiteitsplein 1, B-2610 Antwerpen, Belgium}
\date{\today}

\begin{abstract}
A theoretical investigation of the single particle polariton properties for a microcavity embedding a charged quantum well is presented. The electron gas optical susceptibility is calculated numerically using the method devised by Combescot and Nozi\`eres. The role of many-body effects, such as the Fermi edge singularity and Anderson orthogonality catastrophe, in the polariton formation is elucidated. 
By tuning the light-matter coupling the short time behaviour of the electron gas response function is probed and comparison with earlier results only using the long time response are made.
Various single particle polariton properties such as the Rabi splitting, line shape, Hopfield coefficients and effective mass are discussed. These are experimentally accessible quantities and thus allow for a comparison with the presented theory.
\end{abstract}

\maketitle

\section{Introduction}\label{sec:Introduction}
Microcavity polaritons are bosonic quasiparticles that appear in a semiconductor microcavity with an embedded quantum well (QW) when the cavity is tuned to an excitonic transition in the QW. In the last decade they have emerged as quasiparticles with many favorable properties for studying quantum optics and many-body physics in integrated photonic structures.
Bose-Einstein condensation of polaritons and their superfluid properties have been investigated in detail both experimentally and theoretically \cite{iac_review}.
In these studies, the tunability of the polariton properties is often exploited. For example, the number of quantum wells is varied to change the Rabi frequency; etching, strain and surface acoustic waves have been used to create potentials for the polaritons. 

A tuning parameter that has received relatively little attention is the introduction of charges in the QW that interact with the polaritons. In the case of low charge concentration, trionic bound states exist and a trionic polariton is observed \cite{Rapaport,Bloch}. An experiment with high density modulation doping was performed by Gabbay {\em et al.} \cite{GabbayCohen}. They have shown that strong light-matter coupling is possible, even though neither an excitonic nor a trionic state could be resolved. 
From the theoretical side, the mixed electronic-polaritonic system was proposed to reach superconductivity at higher temperatures, possibly even room temperature, thanks to a strongly attractive electron-electron interaction mediated by the polaritons \cite{laussy}. For what concerns the effect of the electron gas on the polaritons, various contributions both from the experimental \cite{Lagoudakis,Perrin,Das} and theoretical \cite{Kavokin} side have been made. Here, the polariton quasiparticle itself was assumed not to be affected by the electron gas. 

The purpose of this paper is to investigate the effect of the two-dimensional electron gas (2DEG) on the single particle polariton properties. A first theoretical study on the modification of the single particle polariton properties due to the charges inside the quantum well started from an estimate of the oscillator strength computed in the context of the Mahan singularity\cite{glazov,Averkiev}. Indeed, the optical excitation of an electron gas (metal or doped semiconductor) is a central problem in many-body physics that was introduced by Mahan \cite{mahan} and to which Nozi\`eres, De Dominicis \cite{nozieres} and Anderson \cite{Anderson} made seminal contributions. They showed that the absorption becomes singular at the so-called Fermi edge $\omega_T$ with some characteristic powerlaw:
\begin{eqnarray*}
\mathcal{A}(\omega) \sim \frac{1}{|\omega-\omega_T|^\alpha}.
\end{eqnarray*}
This Fermi edge singularity is still an active topic of research that has recently attracted the attention of the community working on ultracold atomic gases \cite{knap,demler,Schirotzek}.  

In this article, we will use a method by Combescot and Nozi\`eres\cite{CN} to calculate fully numerically the electron gas optical susceptibility for all frequencies. Therefore, we also have access to energies far away from the Fermi edge singularity.

The optical modes in a microcavity with an embedded quantum well are given by the photon spectral function $\mathcal{A}(\omega)$, defined as
\begin{eqnarray}
\label{eq:specfunc}
\mathcal{A}(\omega) &=& -\frac{1}{\pi}\textrm{Im} D^{\textrm{ret}}(\omega)\,, \crcr
D(i\omega) &=& \frac{1}{\omega - \omega_c - \mathcal{F}(i\omega)}\,,
\label{eq:AD}
\end{eqnarray}
where $D^{\textrm{ret}}(\omega)=D(i\omega\to\omega+i\eta^+)$ is the retarded photon Green's function with $\omega_c$ the energy of the bare cavity mode (i.e. no quantum well inside). The momentum dependence is neglected, because photon momenta are always negligible as compared to the electron and hole momenta.  Cavity losses can be easily incorporated by giving $\omega_c$ an imaginary part. We prefer not to include them in our analysis, in order to highlight the polariton properties stemming from the electron-hole dynamics.
The photon self energy due to interactions with the quantum well is given by $\mathcal{F}(i\omega)$. The latter quantity is nothing but the optical susceptibility of the 2DEG inside the quantum well. The hybrid light-matter modes are thus found if we succeed in calculating $\mathcal{F}(i\omega)$.

This paper is organized as follows: in Section \ref{sec:Method} the formalism for calculating the electron gas response function and the notations used in this paper are presented. Section \ref{sec:Spectra} discusses the polariton single particle properties such as the polariton energies, linewidth, Hopfield coefficients, effective mass and Rabi splitting. The last section \ref{ConclusionAndOutlook} gives a summary and outlook. 

\section{Optical susceptibility of the 2DEG}\label{sec:Method}
Following Combescot and Nozi\`eres\cite{CN} (CN), the optical response of the 2DEG can be described as a sudden quench of the quiescent Fermi sea of conduction electrons with a scattering potential, induced by the appearance of the valence band hole, at the time of excitation. The transient behaviour of the unperturbed single particle eigenstates in presence of this scattering potential determines the full optical response. Within linear response theory and for a valence band hole with infinite mass, the optical susceptibility is given by
\begin{eqnarray}
\mathcal{F}(t-t^\prime) = -i\sum_{{\bf k,k}^\prime} g^*_{\bf k} g_{\bf{k}^\prime} \langle 0 | \mathcal{T} \hat{b}(t)\hat{c}_{\bf k}(t)\hat{c}^\dagger_{\bf k^\prime}(t^\prime)\hat{b}^\dagger(t^\prime)  |0 \rangle.
\label{eq:Ft}
\end{eqnarray}
Here, $g_{\bf k}= \langle c_{\bf k}|\hat{d}|v\rangle$ is the optical interband matrix element and $\hat{c}^\dagger_{\bf k},\, \hat{b}^\dagger$ are the creation operators for a conduction band electron, resp. valence band hole. $\mathcal{T}$ is the usual time-ordering operator.
The optical susceptibility $\mathcal F(i\omega)$ in Eq. \eqref{eq:specfunc} is given by the Laplace transform of the real time response function:
\begin{eqnarray}
\mathcal{F}(i\omega) = \int_{0}^\infty e^{i \omega t } \mathcal{F}(t) \, dt .
\end{eqnarray}
Upon neglecting electron-electron interactions, we assume for the initial state $|0\rangle$, i.e. before optical excitation, a Slater determinant of single particle plane wave states $|\bf{K}\rangle$. The response function \eqref{eq:Ft} can then be recasted into the form (see Ref. \onlinecite{CN} for details)
\begin{eqnarray}
\mathcal{F}(t) = -iL(t)G(t)
\label{FLG}
\end{eqnarray}
with
\begin{equation}
G(t) = -e^{-i\varepsilon_h t}\textrm{det}[\lambda_{KK^\prime < k_F}(t)]
\label{eq:G}
\end{equation}
\begin{align}
-L(t) = &
\sum_{p,p^\prime > k_F} g_p g^*_{p^\prime} e^{-i\varepsilon_p t} \crcr
  \times & [\lambda_{pp^\prime} - 
\sum_{KK^\prime < k_F}\lambda_{pK}(\lambda^{-1})_{KK^\prime}\lambda_{K^\prime p^\prime} ],
\label{eq:L}
\end{align}
where the key quantity is the time-dependent matrix $\lambda(t)$:
\begin{eqnarray}
\lambda_{KK^\prime}(t) = \sum_{\bar{p}} \langle K|\bar{p}\rangle \langle \bar{p} |K^\prime \rangle  e^{-i(\bar{\varepsilon}_p-\varepsilon_K) t}.
\label{eq:lambda}
\end{eqnarray}
This matrix $\lambda(t)$ contains the overlaps between the single particle plane wave states $|K\rangle$ with energy $\varepsilon_k=k^2/2$ (units are such that $\hbar=m_e=1$ and the bottom of the conduction band is chosen as the zero of energy) before the transition, and the single particle scattering states satisfying $H_f|\bar{p}\rangle=\bar{\varepsilon}_p|\bar{p}\rangle$, where the latter Hamiltonian is given by
\begin{eqnarray}
H_f &=& \varepsilon_h + \sum_k \varepsilon_k \hat{c}^\dagger_k \hat{c}_k + \sum_{k,k^\prime} V_{kk^\prime} \hat{c}^\dagger_{k^\prime}\hat{c}_k. 
\label{Hamiltonian}
\end{eqnarray}
Due to its infinite mass, the valence band hole with energy $\varepsilon_h$ acts as a static external single particle potential for the conduction electrons. In the above we assumed that the photon only couples to zero angular momentum states $l=0$, hence the summation indices in the above expressions are one-dimensional variables. Physically, the separate terms appearing in the expressions \eqref{eq:G} and \eqref{eq:L} have the following interpretation; Eq. \eqref{eq:G} accounts for the dynamical self energy of the valence band hole. The first term in Eq. \eqref{eq:L} represents the direct scattering of an electron above the Fermi sea on the valence band hole potential while the second term denotes the indirect scattering. The latter is mediated by electrons inside the Fermi sea.\cite{PH} 
In the remaining part of this paper we work with a scattering potential that is separable in $k$-space, meaning $V_{kk^\prime}=Vu_ku_{k^\prime}$ with $u_k$ some form factor, which is the same choice as in CN. In particular we focus on the case of an attractive potential $V<0$, for which it is known that in two dimensions this always results in the presence of a bound state. Because of this particular choice we can check our numerical results against the analytical results found in CN.

\begin{figure}[hbtp]
\includegraphics[height=47mm,width=1\columnwidth]{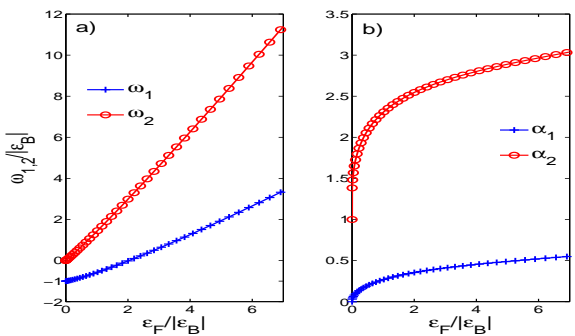}
\caption{a) thresholds $\omega_{1,2}$ normalized to the binding energy of the bound state $\varepsilon_B$  as a function of Fermi energy. b) shows the corresponding powerlaw exponents $\alpha_{1,2}$ as a function of Fermi energy. The exponent $\alpha_1$ is seen to be smaller than one, corresponding to a singular behaviour in absorption. Restoring units, $V=-2.5 \frac{\hbar^2}{m_e}$ was used for the strength of the potential.}
\label{fig:exponentsthresholds}
\end{figure}

In the long time limit $t\gg\varepsilon^{-1}_F$,  CN analytically showed that the response function is given as a sum of two powerlaws:
\begin{equation}
\mathcal{F}(t\gg\varepsilon_F^{-1})  = C_1 \frac{e^{i\omega_1t}}{t^{\alpha_1}}+C_2 \frac{e^{i\omega_2t}}{t^{\alpha_2}}
\label{eq:Fasymp}
\end{equation}
with $C_{1,2}$ some constants. The powerlaw decay is a manifestation of the so-called Anderson orthogonality catastrophe\cite{Anderson} which states that the many-particle wave functions with and without the presence of the scattering potential are orthogonal to each other. The scattering potential creates multiple low-energy intraband electron-hole pairs near the Fermi level and as such, the overlap of the wave function before and after the optical excitation decreases with time as a powerlaw. For the optical absorption, there are two different thresholds:
\begin{eqnarray}
\omega_1 &=& \varepsilon_B+\Delta+\varepsilon_F\,, \crcr
\omega_2 &=& \Delta+2\varepsilon_F.
\label{eq:thresholds}
\end{eqnarray}
In a non-interacting picture, when filling up all single particle energy levels, the minimum energy required for absorption would be the Fermi energy $\varepsilon_F$, i.e. the energy to put an electron into the lowest unoccupied state in the conduction band. However, after the appearance of the scattering potential an energy shift $\Delta = \sum_{p<k_F} (\bar{\varepsilon}_p-\varepsilon_p)$ arises due to the filling of the new single particle energy levels $\bar{\varepsilon}_p$. Finally, also the energy level of the bound state $\varepsilon_B<0$ should be added. This corresponds to the first threshold $\omega_1$. The second threshold $\omega_2$ arises from the breaking up of the bound state; the particle occupying the bound state energy level should be put on top of the Fermi sea. This cost is $|\varepsilon_B|+\varepsilon_F$ and should be added to the first threshold.
The powerlaw exponents $\alpha_{1,2}$ in Eq. \eqref{eq:Fasymp} can be related to the scattering phase shift at the Fermi level of the conduction electrons scattering of the potential\cite{CN}:
\begin{eqnarray}
\alpha_1 &=& \left(\frac{\delta_{k_F}}{\pi}-1\right)^2\,,\crcr
\alpha_2 &=& \left(\frac{\delta_{k_F}}{\pi}-2\right)^2.
\label{eq:PLexponents}
\end{eqnarray}
According to Friedel\cite{friedel}, the phase shift at momentum $p$ is given as
\begin{eqnarray}
\delta_p = (\varepsilon_p-\bar{\varepsilon}_p)\pi\nu_p,
\end{eqnarray} 
with $\nu_p$ the density of states. For the Hamiltonian \eqref{Hamiltonian} and the particular choice of the separable potential, the threshold energies \eqref{eq:thresholds} and the corresponding powerlaw exponents \eqref{eq:PLexponents} are depicted in Fig. \ref{fig:exponentsthresholds} as a a function of Fermi energy.

The exponent of the first threshold $\omega_1$ satisfies $\alpha_1<1$ for all densities, meaning that this first threshold exhibits the so-called `Fermi edge singularity':  the optical absorption exhibits singular behaviour at $\omega_1$, with a  powerlaw tail (the divergence of the optical absorption is readily seen from the Laplace transform of the first term in Eq. \eqref{eq:Fasymp}, which goes like $|\omega-\omega_1|^{\alpha_1-1}$).
For the second exponent $\alpha_2>1$ holds, corresponding to a non-singular onset of absorption.  In the case of an undoped quantum well it holds that $\alpha_1=0$ and the first threshold corresponds to the excitation of the well-known quantum well exciton for which the absorption is most singular, i.e a delta function. The second threshold than corresponds to the onset of continuum absorption, above which free interband electron-hole pairs are excited. 

We have numerically computed the response function $\mathcal{F}(t)$ in the time domain using the Eqs. \eqref{FLG}-\eqref{Hamiltonian} for the separable model potential mentioned earlier. We computed the time series up to $t\simeq 100\varepsilon^{-1}_F$, far in the asymptotic regime. In the long time limit the singular threshold dominates the response function and Eq. \eqref{eq:Fasymp} can than be approximated by its first term:
\begin{eqnarray}
\mathcal{F}_{SPA}(t)  \equiv C_1 \frac{e^{i\omega_1t}}{t^{\alpha_1}}.
\label{eq:FsinglePL}
\end{eqnarray} 
We extrapolate the numerical time series using this expression. To avoid any manifestations of the Gibbs phenomenon in the frequency domain we damp the time evolution with an exponential with a relaxation time $\tau$ much larger than $t\simeq 100\varepsilon^{-1}_F$ such that we do not destroy the important features of the long-time behaviour of the response function. This numerical damping time sets our resolution around threshold; in frequency space we cannot look closer to threshold than approximately $1/\tau$.

\begin{figure}[hbtp]
\includegraphics[height=75mm,width=1\columnwidth]{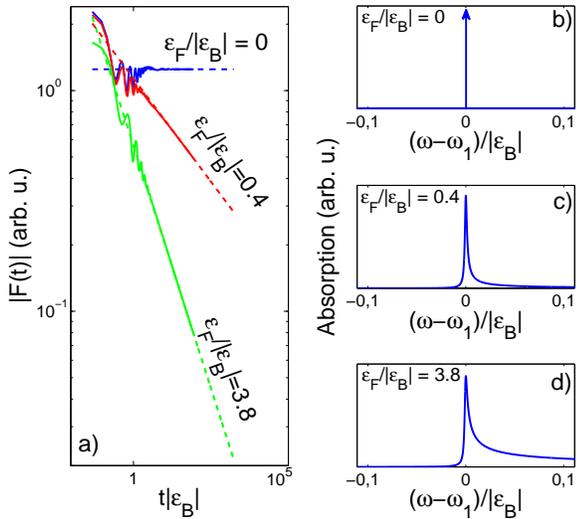} 
\caption{a) modulus of the numerically calculated response function $|\mathcal{F}(t)|$ (solid lines) and the fitted powerlaw (dashed) for several densities as a function of time. (b-d) optical absorption (arb. units) of the electron gas. For zero density, absorption is due to the bound state yielding a delta function. For higher densities the absorption becomes broadened with a large right asymmetric tail.}
\label{fig:FESspectra}
\end{figure}

In Fig. \ref{fig:FESspectra}a the modulus of the response function is plotted on a double logaritmic scale to evidence the powerlaw nature of the decay. Solid lines correspond to the full numerical simulation.
For zero density the response function asymptotically reaches a constant value, while for higher densities the response function decays over several orders of magnitude reflecting the Anderson orthogonality catastrophe. Figs. \ref{fig:FESspectra}(b-d) give the absorption proportional to $-\textrm{Im}[\mathcal{F}(i\omega)]$ for the same densities considered in the left panel. For an undoped quantum well, the absorption becomes a delta function which is typical for the absorption due to a well-defined quasiparticle, in this case an exciton-like state. For increasing electron gas density, the `excitonic' absorption evolves continuously towards a lineshape exhibiting the so-called Fermi edge singularity: a sharp onset and a powerlaw tail on the right hand side of the singularity. Note that in theory the onset should be a sudden step, but due to numerical damping with an exponential it is slightly rounded.

\begin{figure}[hbtp]
\includegraphics[height=50mm,width=0.8\columnwidth]{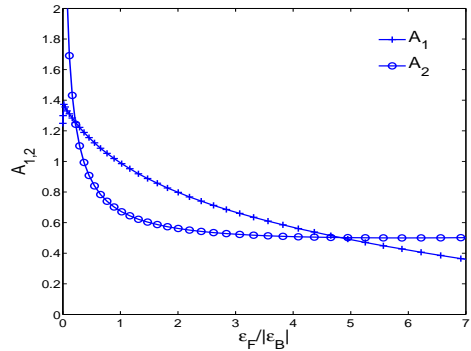} 
\caption{Numerically extracted values for $A_1$ in Eq. \eqref{eq:SPLaplace} (crosses) and $A_2$ in Eq. \eqref{eq:FwGlazov} as found in Ref. \onlinecite{glazov} (circles) as a function of Fermi energy. }
\label{fig:C}
\end{figure}

To extract the constant $C_1$ in Eq. \eqref{eq:FsinglePL}, we performed a numerical fit of the above single powerlaw approximation (SPA) to the long time behaviour of the full numerical result. This is indicated by the dashed lines in Fig. \ref{fig:FESspectra}a. The constant $C_1$ is in general a complex number. However, because the absorption should in theory contain a sharp onset at the threshold energy, this fixes the phase of $C_1$.  The Laplace transform of the above expression \eqref{eq:FsinglePL} then becomes
\begin{eqnarray}
\mathcal{F}_{SPA}(\omega)&=&\frac{A_1e^{-i\frac{\pi\alpha_1}{2}}}{(\omega-\omega_1+i\eta^+)^{1-\alpha_1}},\crcr
A_1 &=& |C_1|\Gamma(1-\alpha_1),
\label{eq:SPLaplace}
\end{eqnarray}
with $\Gamma(x)$ the Gamma function. The fitted value $A_1$ as a function of Fermi energy is depicted in Fig. \ref{fig:C} represented by the crosses. As expected, it is of order one.

At zero density the exact response function is found to be
\begin{eqnarray}
\mathcal{F}_{k_F=0}(t) = -i\sum_n e^{-i\bar{\varepsilon}_nt}|\phi_n(r=0)|^2,
\end{eqnarray}
by using Eqs. (\ref{FLG}-\ref{Hamiltonian}). The Laplace transform is given by
\begin{eqnarray}
\mathcal{F}_{k_F=0}(\omega) = \sum_n \frac{|\phi_n(r=0)|^2}{\omega-\bar{\varepsilon}_n+i\eta^+}.
\label{eq:Ftwolevel}
\end{eqnarray}
In the regime where the SPA is valid, only the bound state $n=0$ determines the optical response. We than see that the SPA becomes exact by comparing Eqs. \eqref{eq:SPLaplace} and \eqref{eq:Ftwolevel} ($\alpha_1=0$ for zero density, see Fig. \ref{fig:exponentsthresholds}b). For the prefactor we find $A_1 = C_1 = |\phi_B(r=0)|^2$ where $\phi_B(r)$ is the relative electron hole wave function of the bound state.

In a previous paper by Averkiev and Glazov \cite{glazov} (AG) the 2DEG optical susceptibility has been calculated, starting from an ad hoc modification of the interband matrix element, mimicking the powerlaw divergence asymptotically close at the first threshold. Their result for the modulus of the 2DEG optical susceptibility is given by
\begin{eqnarray}
 |\mathcal{F}_{AG}(\omega)|&=&\frac{A_2}{|\omega-\omega_1|^{1-\alpha_1}},\crcr
 A_2 &=& \frac{1}{2\sin(\pi(1-\alpha_1))}.
\label{eq:FwGlazov}
\end{eqnarray}

Both Eqs. \eqref{eq:SPLaplace} and \eqref{eq:FwGlazov} have the same functional form. To compare both methods, we have plotted $A_2$ on Fig. \ref{fig:C} with the open circles. For low densities the approximations made in Ref. \cite{glazov} break down, where for higher densities, it reproduces the correct order of magnitude.

\section{Polariton properties} \label{sec:Spectra}
In this paper we are interested in the eigenmodes of the coupled cavity-quantum well system, when a 2DEG is present in the quantum well. The main quantity is now given by the photon spectral function Eq. \eqref{eq:specfunc}, in which we use for the optical susceptibility $\mathcal{F}(i\omega)$ the results from the previous section. 


\subsection{Polariton energies, linewidth}
\begin{figure}[h!]
\includegraphics[scale=0.35]{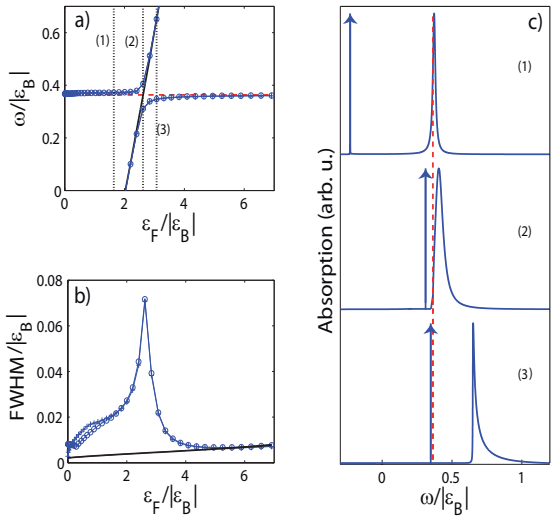}
\caption{a) Avoided crossing between the singular threshold (black solid) and the cavity mode (red dashed). The Fermi energy serves as the detuning parameter while the photon energy (red dashed) is fixed. Black dotted lines indicate several cuts at fixed densities. The corresponding lineshapes of these cuts are depicted in (c) as a function of frequency. There again, the red dashed line is the photon energy. Figure b) shows the upper polariton full-width at half maximum (crosses, circles) as a function of Fermi energy.  A light-matter coupling $g=0.1 \sqrt{|\varepsilon_B|}$ is used. }
\label{fig:PS1}
\end{figure} 

Because the singular threshold $\omega_1$ shifts with the electron density, see Fig. \ref{fig:exponentsthresholds}a, we can keep the cavity mode energy $\omega_c$ constant and use the 2DEG density as the tuning parameter. 
 A typical anti-crossing spectrum for the choice $ g=0.1 \sqrt{|\varepsilon_B|}$ in Eq. \eqref{eq:L} is depicted in Fig. \ref{fig:PS1}a where the cavity mode (red dashed) is fixed. The singular threshold is depicted as the black solid line. Open circles mark the polariton energies, the poles of $\mathcal A(\omega)$, given by Eq. \eqref{eq:AD} using the full numerical time series for the optical susceptibility $\mathcal{F}(\omega)$, computed from Eqns. (\ref{eq:Ft}-\ref{eq:lambda}).
 
Some representative line shapes are shown in Fig.~\ref{fig:PS1}c. Within our treatment, the lower polariton has zero linewidth. This is due to the fact that the lower polariton energy always lies below the threshold energy $\omega_1$, where ${\rm Im} [\mathcal{F}(i \omega)]=0$. At the lower polariton energy, there are no decay channels into matter excitations. In practice, the lower polariton linewidth will thus be determined by the photon life time, at least for the case of the infinite hole mass. When the hole mass is finite, hole relaxation will reduce the life time of the lower polariton \cite{ourpreprint,disorder}
 
The upper polariton on the other hand does show an intrinsic broadening due to the electron gas dynamics. In particular, around resonance, the upper polariton inherits the strong asymmetric form from the electron gas optical absorption. Its full width at half maximum (FWHM) is plotted in Fig. \ref{fig:PS1}b. The FWHM reaches its maximum at zero detuning between the photon and the singular threshold.
At high densities the upper polariton linewidth (open circles) is seen to approach the linewidth of the electron gas response function (black solid line), as expected. The latter increases itself as a function of density, because the exponent $\alpha_1$ of the singular threshold increases as a function of density, see Fig. \ref{fig:exponentsthresholds}b. 

\begin{figure}[h!]
\includegraphics[scale=0.35]{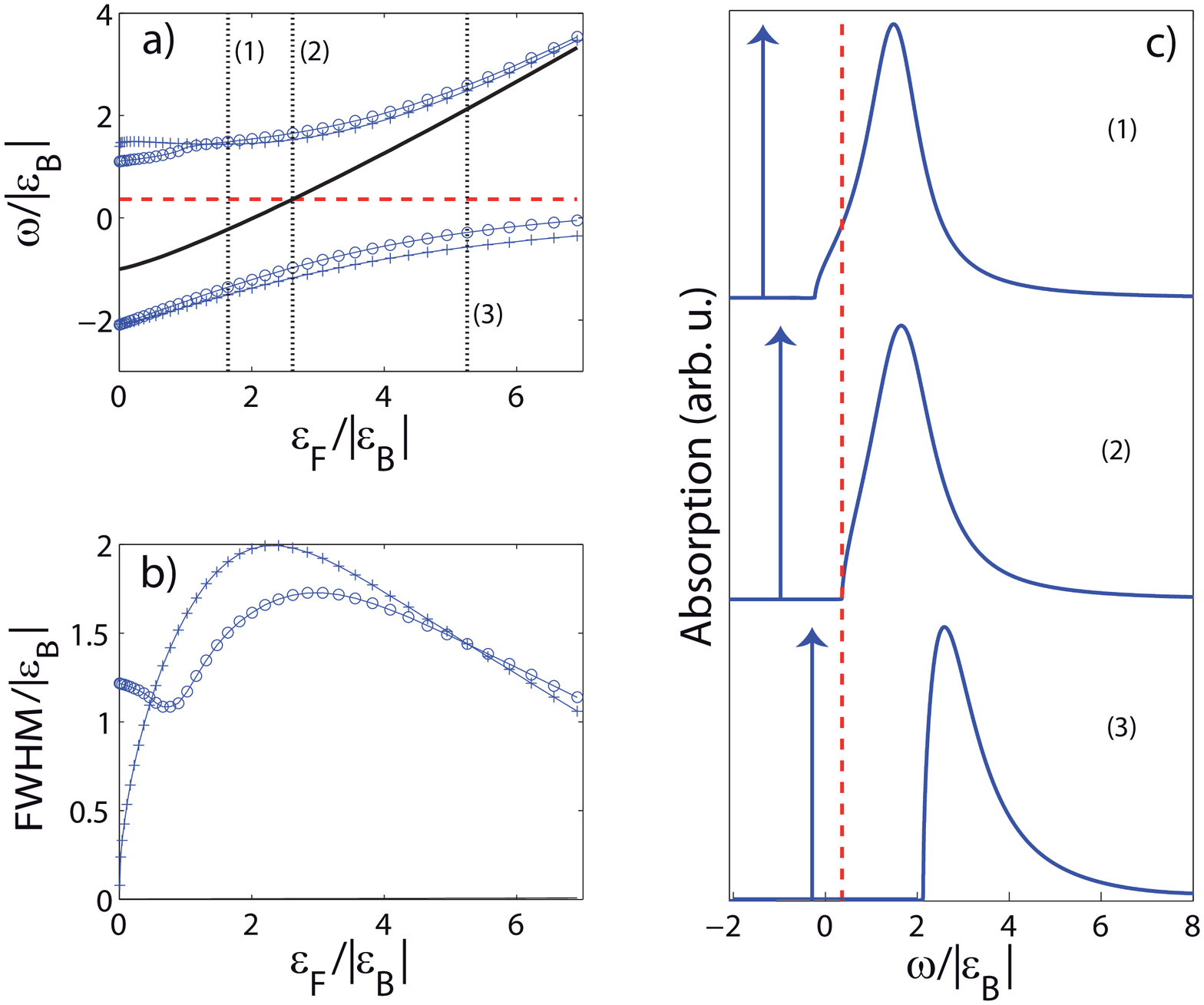}
\caption{(a-c) present the same quantities as in Fig. \ref{fig:PS1}, but for a larger light-matter coupling strength $g=1.5\sqrt{|\varepsilon_B|}$.  The discrepancy between the full numerical result (circles) and the long time response (crosses) appears in Figs. (a-b), because the Rabi oscillations are now faster than the Fermi time.}
\label{fig:PS2}
\end{figure}

The crosses in Figs. \ref{fig:PS1} (a,b) are obtained by using for $\mathcal F(\omega)$ in Eq. \eqref{eq:AD} the asymptotic power-law approximation Eq. \eqref{eq:SPLaplace}. The agreement with the full numerics is very good, except  for a small discrepancy in the upper polariton linewidth at small density.  The reason is that at lower densities the power law approximation becomes less justified. The time scale for the asymptotic regime to set in is determined by the inverse Fermi energy and thus becomes longer for lower densities. The condition for the SPA to work is that the time scales probed by the polariton are longer than the Fermi time. The accuracy of the SPA is therefore expected to become worse for larger light-matter couplings, so that the shorter time optical response of the quantum well is probed. This expectation is borne out in Fig. \ref{fig:PS2}, where we plot the same quantities as in Fig. \ref{fig:PS1}, but for a larger light-matter coupling strength $g=1.5\sqrt{|\varepsilon_B|}$, vs. $g=0.1\sqrt{|\varepsilon_B|}$ in Fig. \ref{fig:PS1}. The discrepancy between the SPA (crosses) and the full numerical result (open circles) is indeed larger. 
 
\subsection{Rabi splitting vs density} \label{subsec:RabiVsDens}
\begin{figure}[htbp]
\includegraphics[height=60mm,width=1\columnwidth]{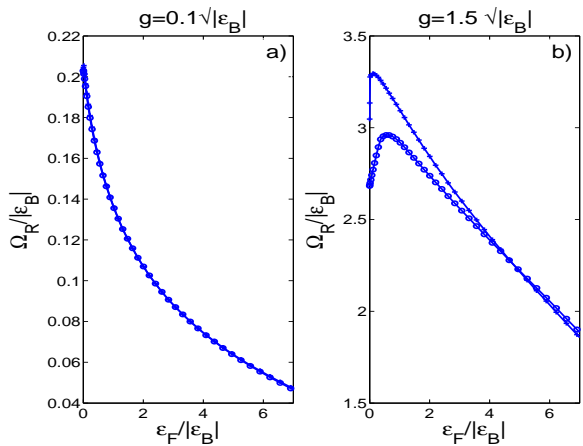}
\caption{Rabi splitting as a function of Fermi energy for two different choices of the light-matter coupling $g$. For larger $g$ a discrepancy between the exact numerical result (open circles) and the SPA (crosses) appears at low densities. Note the different scales on the vertical axes.}
\label{fig:RabiVsDens}
\end{figure}

Instead of fixing the cavity mode, we can tune the photon into resonance with the singular threshold for every density. We then obtain the Rabi splitting as a function of Fermi energy shown in Fig. \ref{fig:RabiVsDens} for two different values of the light-matter coupling $g=0.1 \sqrt{|\varepsilon_B|},\, g=1.5 \sqrt{|\varepsilon_B|}$.  
Note that, similar to Figs. (\ref{fig:PS1}-\ref{fig:PS2}), the difference between the full numerical result (circles) and the SPA (crosses) becomes larger for stronger light-matter coupling and that the discrepancy is largest at low densities.
In general the Rabi splitting decreases as a function of the Fermi energy for high densities. We attribute this to the Anderson Orthogonality Catastrophe; the recombination of all the intraband electron-hole pairs excited due to the scattering potential into a single photon becomes less probable so that the overall light-matter overlap decreases. As more intraband pairs get created for increasing electron density, the overall Rabi splitting decreases as a function of density.

\begin{figure}[hbtp]
\includegraphics[height=50mm,width=0.9\columnwidth]{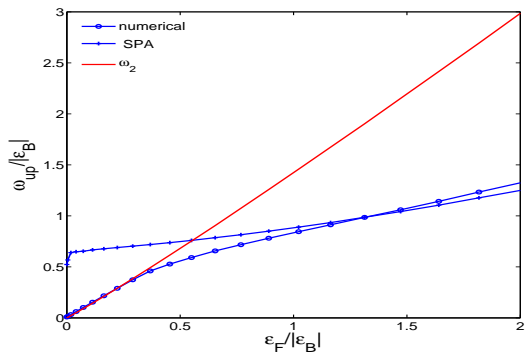} 
\caption{Upper polariton energy $\omega_{\textrm{up}}$ (circles = fully numerical; crosses = SPA) as a function of Fermi energy. Red solid line depicts the second threshold $\omega_2$. $g=1.5 \sqrt{|\varepsilon_B|}$ is used for the light-matter coupling strength. }
\label{fig:secondthreshold}
\end{figure}

Furthermore, we observe a small increase of the Rabi splitting at small Fermi energy in Fig. \ref{fig:RabiVsDens}b. We attribute it to the influence of the continuum (the second threshold $\omega_2$) on the upper polariton energy. This is illustrated in Fig. \ref{fig:secondthreshold}, where we show the upper polariton energy (circles) as a function of Fermi energy. The second threshold $\omega_2$ is depicted by the red solid line. The upper polariton energy is seen to follow $\omega_2$ in the region where the Rabi splitting is increasing (note the different scales on Figs. \ref{fig:RabiVsDens}b and \ref{fig:secondthreshold}). For larger Fermi energy $\omega_2$ is pushed to higher energies, see Eq. \eqref{eq:thresholds}. The influence of the continuum than becomes negligible and our numerical results are close to the SPA (crosses). 

\subsection{Spectral weights}
A further useful characterization of the polariton state is offered by the Hopfield coefficients, that express how much the character of the polariton is photon or matter like. The polariton wave function can be written as
\begin{eqnarray*}
|\Psi_{\textrm {LP,UP}}\rangle  = C_{\textrm {LP,UP}}|\Psi_{\textrm {photon}}\rangle -\sqrt{1-|C_{\textrm {LP,UP}}|^2}|\Psi_{\textrm {matter}}\rangle,
\end{eqnarray*}
defining the photonic Hopfield coefficient $C_{\textrm {LP,UP}}$ of the lower and upper polaritons respectively. In terms of the photonic spectral function, the photonic Hopfield coefficient of the lower polariton is given by the integral of the spectral function around the lower polariton energy. We identify the upper polariton Hopfield coefficient with the integral over the other frequencies. According to the sum rule for the spectral function,
\begin{eqnarray*}
1=\int_{-\infty}^\infty \textrm{d}\omega\, \mathcal{A}(\omega)
\end{eqnarray*}
we then simply have $|C_{\textrm {UP}}|^2=1-|C_{\textrm {LP}}|^2$.

\begin{figure}[htbp]
\includegraphics[height=85mm,width=1\columnwidth]{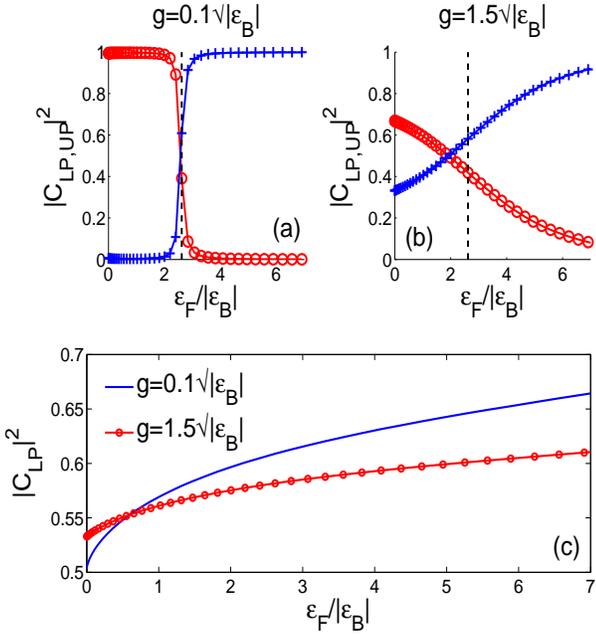}
\caption{(a-b) Photonic Hopfield coefficients for two different strengths of light-matter coupling as a function of Fermi energy. Crosses (circles) represent the photonic Hopfield coefficient for the lower (upper) polariton. The cavity mode was taken the same as in Figs. (\ref{fig:PS1}-\ref{fig:PS2}) c) Lower polariton photonic Hopfield coefficient when the photon is taken into resonance with the singular threshold for all densities.}
\label{fig:hopfields}
\end{figure}

Fig. \ref{fig:hopfields}(a-b) shows the photonic Hopfield coefficients for the lower (blue crosses) and upper (red circles) polariton  as a function of Fermi energy. The fixed cavity mode energy was chosen the same as in Fig. \ref{fig:PS1},\ref{fig:PS2}. The Fermi energy serves again as the detuning parameter and the resonance is represented by the black dashed line. For large Fermi energy, the photon is strongly red-detuned with respect to the singular threshold and the lower polariton becomes almost purely photonic. As in the case of the simple exciton-polariton, the character of the lower polariton crosses over from matter like at positive photon detuning to photon like at negative photon detuning for small light-matter coupling. The detuning window in which the crossover takes place, is determined by the Rabi frequency. 

Remarkably, for larger light-matter coupling the Hopfield coefficients at resonance are different from the value $1/2$, that is obtained for a simple two-level system. 
The photonic content of the lower polariton is plotted as a function of Fermi energy in Fig. \ref{fig:hopfields}c, where the photon was taken into resonance with the singular threshold for all Fermi energies. Only at zero density and small light-matter coupling, the polariton consists of equal amounts of photon and matter. For larger light-matter coupling, $g=1.5 \sqrt{|\varepsilon_B|}$ vs $g=0.1 \sqrt{|\varepsilon_B|}$, not only the bound state contributes to the optical response but also the continuum states become an important part of the description, resulting in $|C_{\textrm{LP}}|^2>1/2$ at zero density.

\subsection{Lower polariton effective mass} \label{subsec:EffMass}
\begin{figure}[h!]
\includegraphics[height=60mm,width=1\columnwidth]{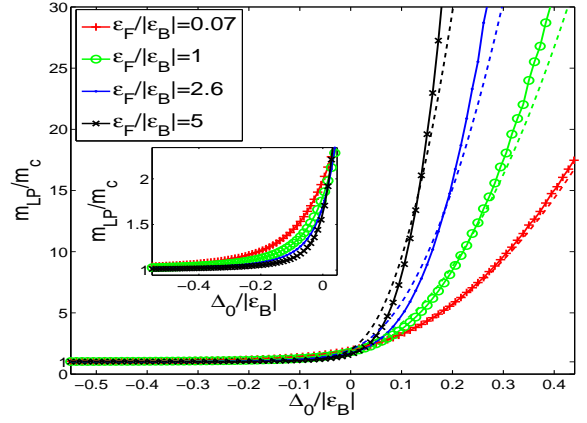}
\caption{Effective mass of the lower polariton (in units of effective photon mass) as a function of $\Delta_0$. Different lines correspond to different electron densities. The inset is a detailed zoom of the left part of the main figure. $g=0.1 \sqrt{|\varepsilon_B|}$ is used for the light-matter coupling strength. The effective mass obtained by using a simple two-level model is depicted by the dashed lines.}
\label{fig:EffectiveMass}
\end{figure}

Within our approximations (in particular the infinite hole mass) the lower polariton is still a quasi particle with infinite intrinsic life time, even in the case of a highly doped quantum well. Its lineshape remains a delta function, despite the electron gas dynamics. The latter only affects the upper polariton lineshape as discussed in previous section. It is therefore interesting to consider the lower polariton effective mass $m_{\textrm{LP}}$. Let's assume a quadratic in-plane dispersion relation for both the cavity photon and lower polariton. Furthermore, we assume that the electron dynamics happens on a length scale much shorter than the polariton wave length (this amounts to assuming a flat dispersion relation for the electron degrees of freedom as compared to the photon dispersion relation).  The lower polariton effective mass is then given by
\begin{equation}
\frac{m_{\textrm{LP}}}{m_{\textrm{c}}}=\left(\frac{\textrm{d} \varepsilon_{\textrm{LP}}}{\textrm{d} \Delta_{0}}\right)^{-1},
\label{eq:mass}
\end{equation}
where $\varepsilon_{LP}$ is the lower polariton energy and $\Delta_{0} = \omega_{\textrm{c}}(k_{\parallel}=0)-\omega_{1}(k_\parallel=0)$ is the detuning at normal incidence (i.e. at zero in-plane momentum $k_\parallel$) between the cavity mode and the singular threshold. In Fig. \ref{fig:EffectiveMass} the lower polariton effective mass as a function of $\Delta_0$ is depicted for several 2DEG densities. For a fixed large $\Delta_0\gg 0$, the higher electron densities yield a larger lower polariton mass. On the other hand, it turns out that in the regime $\Delta_0\lesssim0$ the behaviour is opposite, as can be seen from the inset in Fig. \ref{fig:EffectiveMass}, showing a more detailed zoom of the left side of the larger figure. There it is seen that the lower polariton mass converges faster towards the bare photon mass for a higher density 2DEG. Both are consequences of the decreased Rabi splitting for a higher density electron gas.

It is instructive to compare our numerical results with the simple two-level model for exciton-polariton formation in an empty quantum well. Then, the lower polariton energy is given by
\begin{eqnarray}
\varepsilon_{\textrm{LP}}=\frac{\omega_c+\omega_1}{2} - \frac{\sqrt{\Delta_0^2+\Omega^2_R}}{2},
\label{eq:twolevel}
\end{eqnarray}
with $\omega_{\textrm{c,(1)}}$ the photon (singular threshold) energy. Combining both Eqs. \eqref{eq:mass}, \eqref{eq:twolevel} we find 
\begin{eqnarray}
\left(\frac{m_{\textrm{LP}}}{m_{\textrm{c}}}\right)_{\textrm{two level}} =\frac{2}{1-\frac{\Delta_0}{\sqrt{\Delta_0^2+\Omega_R^2}}}.
\label{eq:twolevelmass}
\end{eqnarray}
In Fig. \ref{fig:EffectiveMass} the dashed lines represent the lower polariton effective mass obtained by using the above two-level model. The Rabi frequency used in the expression \eqref{eq:twolevelmass} is taken from Fig. \ref{fig:RabiVsDens}a. In the limit of zero electron density the two-level model coincides with the full numerical result. This is expected since for $\Omega_R \ll |\varepsilon_B|$ we are only probing the long time response of the electron gas (for which the SPA becomes exact at zero density). For nonzero electron densities, a difference between the SPA and the two-level system becomes apparent at positive $\Delta_0$: due to the presence of the electron gas, the lower polariton appears to be heavier than expected on the basis of the two-level model.

\section{Conclusion and outlook}\label{ConclusionAndOutlook}
We have discussed the linear properties of the hybrid light-matter states occurring as the result of the strong coupling between a microcavity photon and an embedded charged quantum well. Within our approximations, it is found that in the presence of the electron gas the lower polariton remains a good quasi particle. We calculated some experimentally accessible quantities such as the photonic Hopfield coefficient and the lower polariton effective mass as a function of 2DEG density. Furthermore, the role of the Fermi edge singularity and Anderson Orthogonality Catastrophe on the upper polariton lineshape and FWHM were elucidated. 
Also, we have shown that the short time dynamics of the electron gas can be probed by increasing the strength of the light-matter coupling.

For what concerns the zero linewidth of the lower polariton, corrections due to a finite hole mass are expected, but its description needs an extension of the CN approach\cite{PH,US,Rosch}.
Also the effect of the spin degree of freedom of the optically injected electron needs further investigation. In the present treatment the spin polarization was the same as the spin of the electrons in the Fermi sea. 

\section{Acknowledgements}

We thank Atac Imamoglu for fruitful discussions. We acknowledge financial support from the UA-BOF `lanceringsproject' and the FWO Odysseus program.


\begin{thebibliography}{99}
\bibitem{iac_review} I. Carusotto, and C. Ciuti, Rev. Mod. Phys. {\bf 85}, 299-366 (2013).
\bibitem{Rapaport} R. Rapaport, R. Harel, E. Cohen, Arza Ron, E. Linder, and L. N.
Pfeiffer, Phys. Rev. Lett. {\bf 84}, 1607 (2000).
\bibitem{Bloch} D. Bajoni, M. Perrin, P. Senellart, A. Lema\^itre, B. Sermage, and J. Bloch, Phys. Rev. B {\bf 73}, 205344 (2006).
\bibitem{GabbayCohen} A. Gabbay, Y. Preezant, E. Cohen, B. M. Ashkinadze, and L. N. Pfeiffer, Phys. Rev. Lett. {\bf 99}, 157402 (2007).
\bibitem{laussy} F. P. Laussy, A. V. Kavokin, and I. A. Shelykh, Phys. Rev. Lett. {\bf 104}, 106402 (2010). 
\bibitem{Lagoudakis} P. G. Lagoudakis, M. D. Martin, J. J. Baumberg, A. Qarry, E. Cohen, and L. N. Pfeiffer, Phys. Rev. Lett. {\bf 90}, 206401 (2003).
\bibitem{Perrin} M. Perrin, P. Senellart, A. Lema\^itre, and J. Bloch, Phys. Rev. B {\bf 72}, 075340 (2005).
\bibitem{Das} A. Das, B. Xiao, S. Bhowmick, and P. Bhattacharya, Applied Physics Letters {\bf 101}, 131112 (2012).
\bibitem{Kavokin} G. Malpuech, A. Kavokin, A. Di Carlo, and J. J. Baumberg, Phys. Rev. B {\bf 65}, 153310 (2002).
\bibitem{glazov} N. S. Averkiev, and M. M. Glazov, Phys. Rev. B {\bf 76}, 045320 (2007).
\bibitem{Averkiev} N. S. Averkiev, M. M. Glazov, and M. M. Voronov, Solid. State. Comm. {\bf 152}, 395 (2012).
\bibitem{mahan} G. D. Mahan, Phys. Rev. {\bf 153}, 882 (1967).
\bibitem{nozieres} P. Nozi\`eres, and C.T. De Dominicis, Phys. Rev. {\bf 178}, 1097 (1969).
\bibitem{Anderson} P. Anderson, Phys. Rev. Lett. {\bf 18}, 1049 (1967).
\bibitem{knap} M. Knap, A. Shashi, Y. Nishida, A. Imambekov, D. A. Abanin, and E. Demler, Phys. Rev. X {\bf 2}, 041020 (2012).
\bibitem{demler} G. Refael, and E. Demler, Phys. Rev. B {\bf 77}, 144511 (2008).
\bibitem{Schirotzek} A. Schirotzek, C. H. Wu, A. Sommer, and M. W. Zwierlein, Phys. Rev. Lett. {\bf 102}, 230402 (2009).
\bibitem{CN} M. Combescot, and P. Nozi\` eres, Le Journal de Physique, {\bf 32}, 913 (1971).
\bibitem{PH} P. Hawrylak, Phys. Rev. B, {\bf 44}, 3821 (1991).
\bibitem{friedel} J. Friedel, Comments on Solid State Phys. {\bf 2}, 40 (1969).
\bibitem{ourpreprint} M. Baeten and M. Wouters,  arXiv:1301.4119

\bibitem{disorder} Neither do we consider effects of inhomogeneous broadening of the quantum well, that will also contribute to the polariton linewidth, but this goes far beyond the scope of the present treatment.


\bibitem{US} T. Uenoyama, and L. J. Sham, Phys. Rev. Lett. {\bf 65}, 1048 (1990).

\bibitem{Rosch} A. Rosch, and T. Kopp, Phys. Rev. Lett. {\bf 75}, 1988 (1995).



\end{thebibliography}
\end{document}